\documentclass[%
 reprint,
 amsmath,amssymb,
 aps,
prb,
]{revtex4-2}

\usepackage{graphicx}
\usepackage{dcolumn}
\usepackage{bm}

\begin{document}


\title{H$_2$ superglass on an amorphous carbon substrate}

\author{M. C. Gordillo$^*$}
\affiliation{Departamento de Sistemas F\'{\i}sicos, Qu\'{\i}micos
y Naturales, Universidad Pablo de
Olavide, Carretera de Utrera km 1, E-41013 Sevilla, Spain}
\affiliation{Instituto Carlos I de Física Teórica y Computacional, Universidad de Granada, E-18071 Granada, Spain}

\author{J. Boronat}
\affiliation{Departament de F\'{\i}sica,
Universitat Polit\`ecnica de Catalunya,
Campus Nord B4-B5, 08034 Barcelona, Spain}

\date{\today}

\begin{abstract}
The phase diagram of a $para$-H$_2$ monolayer absorbed on a experimentally
syntetized amorphous carbon sheet was calculated
using a diffusion Monte Carlo technique. We found that the ground state of
that system changed drastically from a perfectly flat substrate to a situation 
in which the carbon atoms were allowed a certain degree of disorder in the $z$ direction.   
In the first case, at zero pressure we have a glass of density 0.056 $\pm$ 0.003 \AA$^{-2}$ in
equilibrium with an incommensurate solid of 0.068 $\pm$ 0.002 \AA$^{-2}$.
At the equilibrium density, the glass was found to have a tiny, but non-negligible
superfluid faction of less than 1 \% (0.44 $\pm$ 0.05 \%). In the 
$z$-disordered substrate, we observe a significant enhancement of the superfluid 
fraction in the glass phase as well as a smaller but not zero value in the 
incommensurate crystal. 
 \end{abstract}

\maketitle


It is well-known that the most stable form of carbon is graphite. 
It is also well-known that one can isolate one of those single carbon layers and obtain a stable structure termed graphene 
\cite{graph1,graph2}. Even tough the electric properties of graphene are quite different from those of a three-dimensional arrangement \cite{graph3,graph4}, theoretical calculations
failed to find any significant difference between the adsorption behavior of 
quantum species ($^4$He, H$_2$ and D$_2$) on graphene and graphite 
\cite{prl2009,prb2010,prbcarmen}. 

\begin{figure}
\begin{center}
\includegraphics[width=0.8\linewidth]{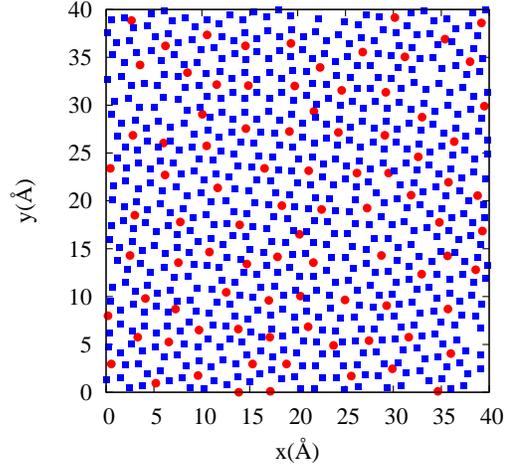}
\caption{Reconstruction of a monolayer amorphous carbon layer as given in Ref. \onlinecite{e3}. Blue squares, carbon atoms. Full red circles, 
adsorption positions for H$_2$ molecules at the equilibrium density of the glass for a planar surface. 
}
\label{cell}
\end{center}
\end{figure}

The honeycomb structure of graphene is made up exclusively of 
carbon hexagons, apart from  occasional defects.  
However, 
amorphous structures, in which we can have carbon pentagons, hexagons and even 
squares in addition
to six-fold rings, can be created by bombarding graphene with an electron beam 
\cite{e1,e2} or synthetized directly   
by chemical vapor deposition \cite{e3}. 
The main features of the latter structure can be captured by a two-dimensional 
40 $\times$ 40 \AA$^2$ patch (Supplementary information of Ref. \onlinecite{e3}) with no holes. 
The projection of those carbon coordinates in the  $x-y$ plare are displayed in Fig. \ref{cell} as 
blue squares. 

The goal of this work is to study the behavior of H$_2$ when adsorbed on an amorphous carbon surface. 
To do so, we will consider that substrate as adequately represented by the the above coordinates, 
but bearing in mind that the carbon layer is not perfectly flat \cite{e3}. We 
solved the
Sch\"odinger equation that describes 
the set of H$_2$ molecules on this new adsorbent using the diffusion Monte Carlo (DMC) method both in flat and corrugated carbon structures. 
Our results show that a stable H$_2$ glass phase is formed  
irrespectively of the substrate. That glass has a tiny superfluid fraction if the underlying carbon sheet is flat, fraction that is considerably
enhanced for the $z$-disordered structure, i.e., we have an stable superglass. 
In the case of H$_2$, there is only a previous theoretical work that predicts a metastable three-dimensional superglass \cite{osychenko}. 
That glass would present a sizeable superfluid density around $\sim$ 1 K. 
                 
The DMC method allows for obtaining exactly the ground state of an ensemble 
of interacting bosons,  within the statistical uncertainties inherent to any 
Monte Carlo technique  
\cite{borobook}. To do so, we have first to write down the 
Hamiltonian describing a monolayer of hydrogen on top of the amorphous carbon substrate. This is: 
\begin{equation} \label{hamiltonian}
  H = \sum_{i=1}^N  \left[ -\frac{\hbar^2}{2m} \nabla_i^2 +
V_{{\rm ext}}(x_i,y_i,z_i) \right] + \sum_{i<j}^N V_{\rm{H_2-H_2}}
(r_{ij}) \ .
\end{equation}
$x_i$, $y_i$, and $z_i$ are the coordinates of the each of the $N$ H$_2$ molecules with mass $m$.
$V_{{\rm ext}}(x_i,y_i,z_i)$ is the interaction potential between each molecule and 
all the carbon atoms in the 40 $\times$ 40 \AA$^2$ patch that models the amorphous structure.  
As in previous works \cite{prb2010,prb2013,prb2022}, that interaction was chosen to be of the Lennard-Jones type, with parameters obtained from Ref. \onlinecite{coleh2}. 
$V_{\rm{H_2-H_2}}$ is modeled by
the standard Silvera and Goldman potential \cite{silvera}. 
As indicated above, we consider two possibilities for the carbon substrate: a flat one, in which the carbon atoms 
are located in the $z =0$ plane, and an irregular one, in which each $z$ 
coordinate was chosen randomly in the interval [-0.4,0.4] \AA.  
This $z$-displacement is 
similar to the vertical distortion of the lattice found in previous ab initio calculations 
of amorphous  
graphite ~\cite{pucker,amorphousg}.           
To avoid the effects in the phase diagram of any particular $z$ carbon 
distribution, all the simulations were repeated ten times with different carbon 
configurations and the results averaged over. 

To actually solve the Schr\"{o}dinger equation defined by the 
many-body Hamiltonian in Eq. \ref{hamiltonian}, 
one uses a trial wave function to reduce the variance to a manageable level.
We use a symmetrized Nosanow-Jastrow wave function split as the product of 
two terms, the first one being: 
\begin{equation}
\Phi_J({\bf r}_1,\ldots,{\bf r}_N) = \prod_{i<j}^{N} \exp \left[-\frac{1}{2}
\left(\frac{b}{r_{ij}} \right)^5 \right] \,
\label{sverlet}
\end{equation}
that depends on the distances, $r_{ij}$, between each pair of H$_2$ molecules and on $b$, a variationally optimized parameter whose
value was found to be 3.195 \AA$ $\cite{prb2010,prb2022}. The second one is: 
\begin{eqnarray}
\Phi_s({\bf r}_{1},\ldots,{\bf r}_{N})  =
\prod_i^{N}  \prod_J^{N_C} \exp \left[ -\frac{1}{2} \left( \frac{b_{{\text
C}}}{r_{iJ}} \right)^5 \right] \nonumber \\
 \times \prod_{I=1}^{N} \left[ \sum_{i=1}^{N} \exp
\{-c [(x_i-x_{\text{site},I})^2 +
(y_i-y_{\text{site},I})^2] \} \ \right] \nonumber \\
\times \prod_i^{N} \exp (-a (z_i-z_{site})^2) \ \ \ \ \   \ 
\label{t2}
\end{eqnarray}
Here, $b_C$ was chosen to be 2.3 \AA, as in previous works 
\cite{prb2010,prb2022}. The $z_{site}$ and $a$ values that minimize 
the energy in  
the infinite dilution limit were $z_{site}$ = 2.94 \AA and $a$=3.06 \AA$^{-1}$.  
If we consider the H$_2$ phase to be 
translationally invariant $c$ = 0, otherwise (i.e. for a solid or glass), $c$ = 0.61 \AA$^{-2}$. 
The latter value for $c$  was taken from Ref. \onlinecite{prb2010} in 
which it was variationally optimized for a incommensurate solid;
nevertheless, we checked that changes in its value of up 50 \% produced 
always worse energies when used  in DMC.  
For both values of $c$ the form of the {\em trial} function allows the H$_2$ molecules 
to be involved in exchanges and recover indistinguishability, something necessary if we are to consider the possibility of a stable superfluid.    
The same form of the trial function was used both for the flat and corrugated carbon substrates.    

\begin{figure}
\begin{center}
\includegraphics[width=0.8\linewidth]{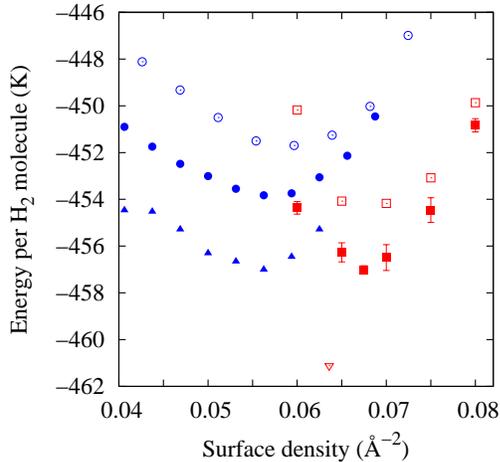}
\caption{Energy per H$_2$ molecule as a function of the density for hydrogen on top of flat graphene (open symbols) and amorphous carbon (full symbols). Circles, quasi two-dimensional liquid; squares, incommensurate triangular solid; full triangles, glass structure; open triangle, $\sqrt 3 \times \sqrt 3$ structure on graphene. When not shown, error bars are of the size of the symbols.    
}
\label{h2conglass}
\end{center}
\end{figure}

In Eq. \ref{t2}, ($x_{\text{site}},y_{\text{site}}$) are the positions of the nodes that define the 
network we are interested in. For a incommensurate hydrogen solid, 
those will be the coordinates of the crystallographic sites of the  
quasi-two dimensional triangular lattice. On the other hand, the glass is defined by a set of 
local energy minima irregularly arranged. To define those minima, we created a two-dimensional grid of  
regularly spaced points at a distance $z_{site}$ above the carbon layer and calculated $V_{{\rm ext}}(x,y,z_{site})$ at such positions. 
After that, we chose the point 
of the grid for which $V_{{\rm ext}}$ is minimum. 
Then, we searched for the point in the grid 
with the next-to-minimum value of the external potential 
located at a distance from the first of at least 
$\sigma_{C-H_2}$ 
(Lennard-Jones parameter of the C-H$_2$ interaction, 2.97 \AA$ $ \cite{coleh2}). This is done to avoid H$_2$-H$_2$ interactions that would contribute 
with positive terms to the total energy in the full DMC scheme. 
The entire process is iterated until it is not possible to locate more hydrogen molecules at distances of at least $\sigma_{C-H_2}$ from each other.
After that, we are left with a  list of ($x_{\text{site}},y_{\text{site}}$) positions ordered from minimum to maximum {\em potential} energy. However, what we need is to 
have a list of nodes ordered from lowest to highest {\em total} energy. To get it, we performed DMC calculations including one single molecule on 
each of those sites and reordered the list with respect to those total single-molecule energies. By following that procedure we minimize the risk of 
getting metastable states once we start filling that network with H$_2$. This 
is so because the difference between the full DMC energy of a system
of $N$ molecules and the set of $N$ increasing individual energies provided in 
the algorithm just described comes from the H$_2$-H$_2$ interaction. 
This contribution relays less and less on the details of the network for 
increasing density, since it depends primarily on the 
average first-neighbor distances. The number of maximum nodes found using this procedure was 104, and that was the maximum number of molecules used to describe
the liquid and glass phases for the density range displayed in Fig. \ref{h2conglass}. 
On the other hand, that number oscillated between 90 and 120 for the incommensurate solid, the number of walkers in the DMC procedure being 300. The remaining of the simulation details are similar
to those in Ref. \onlinecite{prb2022} and omitted here for simplicity.     
The locations of the nodes of the glass are displayed in Fig. \ref{cell} as red circles on top of the carbon 
coordinates (blue squares). To be sure that the election of the cutoff distance does not change the nodes of the glass network, we repeated the entire procedure for 
exclusion values $\sigma_{C-H_2} \pm$ 10\%, finding exactly the same positions for the minima. We checked also that for the densities considered 
in Fig. \ref{h2conglass} to fill the glass network in a different order that 
the one  described above, or to consider a different set of nodes (by starting the building up from another node), did not 
alter the total energies in the density range displayed there.     


In Fig. \ref{h2conglass}, we show 
the energy per H$_2$ molecule as a function of the two-dimensional density for the three phases 
considered in this work: a liquid (full circles), an incommensurate triangular solid (full squares), and a glass (full triangles) 
on a carbon flat amorphous substrate. 
In that figure, we display also the results for graphene, taken from Ref. \cite{prb2010}. Since both the graphene and the amorphous substrate have the same carbon density,
0.38 \AA$^{-2}$, this will allow us to assess the effects of the randomness on the phase diagram of the two-dimensional H$_2$.  What we see is that, at least in this case,
the disorder in the substrate makes both the liquid and solid phases more stable than their corresponding counterparts in graphene. 
In any case, the triangular solid is still more stable that the liquid by 3.1 K at the densities corresponding to zero pressure 
(liquid binding energy, 453.8 $\pm$ 0.5 K; solid binding energy 456.9 $\pm$ 0.5 K).    
Obviously, the lack of periodicity makes impossible to have a commensurate structure, its place being taken by a glass arrangement of variable density. According to Fig. \ref{h2conglass},
the maximum binding energy for this structure is 457.0 $\pm$ 0.5 K at a density of $\rho$= 0.056 $\pm$ 0.003 \AA$^{-2}$. This density is appreciably smaller than the 0.068 $\pm$ 0.002 
\AA$^{-2}$ corresponding to the solid at the minimum of its curve, but equal to the one corresponding to the liquid structure ($\rho$= 0.057 $\pm$ 0.003 \AA$^{-2}$). However, the 
irregularity of the substrate produces a less stable phase than the $\sqrt 3 \times \sqrt 3$ solid in graphene.   
In any case, from the results displayed in Fig. \ref{h2conglass} we can draw an horizontal  double-tangent Maxwell construction line between the minima of the glass and solid curves.
This means that between 0.056 and 0.068 \AA$^{-2}$, we would have a mixture of a glass and a
triangular solid in the adequate proportions to produce a system with the 
desired density. From $0.068 \pm 0.002$ \AA$^{-2}$ up, the stable phase will be a 
triangular solid. 
  
\begin{figure}
\begin{center}
\includegraphics[width=0.8\linewidth]{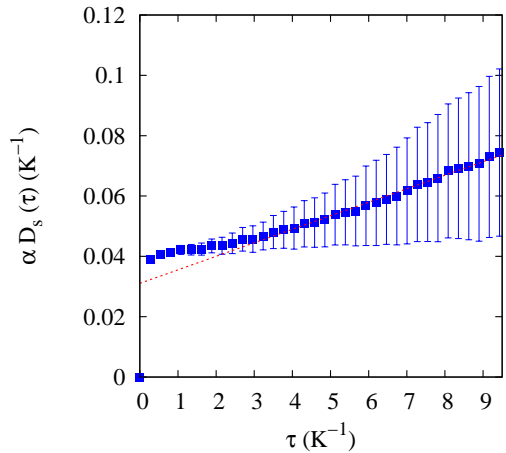}
\caption{Estimator of the superfluid
density for the glass phase at its equilibrium density
Full squares, simulation results.
The straight line represent a linear least-squares fit to the symbols displayed for 
$ \tau >$  3 K$^{-1}$. Since the slope is different from zero, the disordered structure is a superglass.   
}
\label{supersolid2}
\end{center}
\end{figure}

A very recent calculation \cite{prb2022} suggests that we can find supersolid 
behavior for a H$_2$ second layer adsorbed on graphite in a very narrow density 
window 
around 0.1650 \AA$^{-2}$. By a supersolid we mean a solid structure 
(diagonal order) with a superfluid fraction different from zero (off-diagonal 
long-range order). By extension, 
a superglass would be a phase in which the molecules are 
arranged in an amorphous setup with a superfluid fraction larger than zero,
Following the same procedure as in that work
we estimated that fraction, $\rho_s/\rho$, both for the equilibrium densities of the glass and incommensurate triangular solids. 
To do so, we used, as in previous literature for similar systems \cite{prl2020,prb2022} the zero-temperature winding number estimator derived in Ref. \onlinecite{gubernatis},
\begin{equation} \label{super}
\frac{\rho_s}{\rho}= \lim_{\tau \to \infty} \alpha \left(
\frac{D_s(\tau)}{\tau} \right) \ ,
\end{equation}
with $\tau$ the imaginary time used in the quantum Monte Carlo
simulation. Here, $\alpha = N_2/(4 D_0)$, $D_0 = \hbar^2/(2m)$, and
$D_s(\tau) = \langle [{\bf R}_{CM}(\tau)-{\bf R}_{CM}(0)]^2 \rangle$. ${\bf
R}_{CM}$ is the
position of the center of mass of the $N$ H$_2$ molecules considering only their $x$ and $y$ coordinates.
The results are shown in Fig. \ref{supersolid2} for the glass phase.  
Each symbol correspond to an average of ten independent Monte Carlo histories for each value of imaginary time, the straight line being a least-squares fit   
to those points. The error bars correspond to the statistical noise.   
The superfluid fraction is the slope of the curve in the limit $\tau \to \infty$. In Fig. \ref{supersolid2} we represent the that value 
instead of the equivalent average of $\alpha D_s(\tau)/\tau$ for each value of $\tau$ because in that way is easier to appreciate the superfluid fraction when its value 
is very small. 
The slope for the glass implies $\rho_s/\rho$ = 0.44 $\pm$ 0.05 \%, of the same order as the result in the second layer of graphite. To increase the number of Monte Carlo 
histories does not change the superfluid fraction within the error bar given for that magnitude.  
The corresponding curve for the incommensurate solid, not shown for simplicity, 
is completely flat, indicating a normal solid.  

\begin{figure}
\begin{center}
\includegraphics[width=0.8\linewidth]{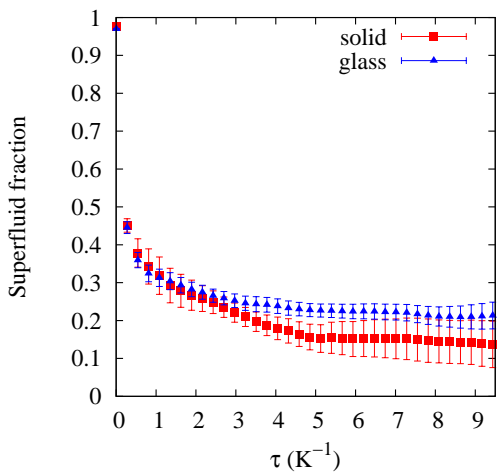}
\caption{Superfluid fraction for the irregular substrate. for two different phases and densities.  Full triangles, glass phase of density $\rho$= 0.053 \AA$^{-2}$;
full squares, triangular solid with $\rho$=0.065 \AA$^{-2}$. 
}
\label{superglass}
\end{center}
\end{figure}

\begin{figure*}
\begin{center}
\includegraphics[width=1.0\linewidth]{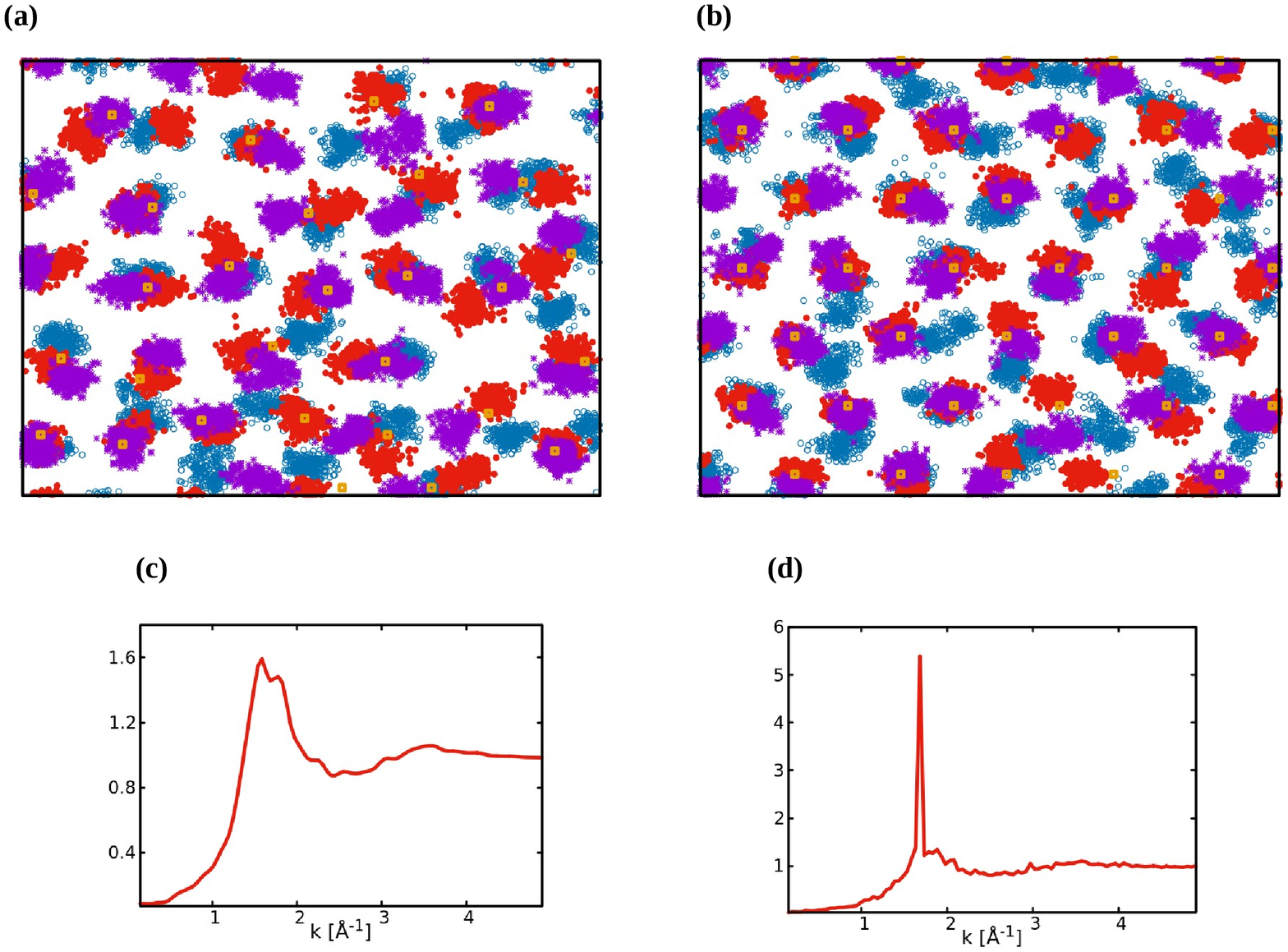}
\caption{ \textbf{(a)} Snapshots representing different sets of $x-y$ configurations corresponding to different Monte Carlo steps for a glass phase of density 0.055 \AA$^{-2}$.  
Different colors stand for different simulation steps. The yellow squares represent the  positions of the nodes of the glass network. 
\textbf{(b)} Same for an incommensurate solid of density 0.065 \AA$^{-2}$.
Now the squares correspond to the crystallographic positions of a triangular lattice. 
\textbf{(c)}-\textbf{(d)}: Static structure factors for the glass and 
incommensurate crystal, respectively. 
}
\label{snaps}
\end{center}
\end{figure*}

Since the amorphous carbon layer is not flat \cite{e3}, we introduced some disorder in the $z$-direction to assess the effects of 
that randomness in the calculated observables. The results for the energies show 
again the same two stable structures.
A double-tangent Maxwell construction indicates a first-order phase transition 
between a glass of density 0.055 $\pm$ 0.003 \AA$^{-2}$, and a two-dimensional 
incommensurate crystal with $\rho$ = 0.0650 $\pm$ 0.0025 \AA$^{-2}$. This means 
that the locus of the coexistence region is basically untouched by the 
introduction of disorder in $z$. However, the change in the superfluid character of the two phases is much relevant.
The results obtained are shown in   Fig. \ref{superglass}.
This figure is similar to Fig. \ref{supersolid2} but, instead of depicting the movement of 
the center of mass, it shows the full superfluid estimator as defined in Eq. \ref{super}. 
The values represented are $\rho_s/\rho$ = 0.21 $\pm$ 0.05 for a glass of 
density 0.053 \AA$^{-2}$ (upper triangles), and $\rho_s/\rho$ = 0.14 $\pm$ 0.05 for a triangular solid with $\rho$ = 0.065 \AA$^{-2}$.  Therefore, our results show that we should have a superglass around 
a density around 0.055 \AA$^{-2}$, independently of the disorder of the substrate in the $z$ direction. Moreover, the disorder in $z$ induces supersolidity also in the incommensurate solid phase, in contrast with the flat adsorption surface.

The finite value of the superfluid fraction in both phases means that particles do not remain isolated around the lattice points but interchanges are possible. 
To show how this feature is observed in the DMC simulations, we plot in Fig.~\ref{snaps} some snapshots for both the glass and incommensurate crystal for the $z$-disordered carbon 
substrate.  Different colors stand for different set of walkers (particle configurations) corresponding to different Monte Carlo steps along the simulation. 
The spreading of every cloud is an indication of the quantum delocalization of the particles. 
One can see that these clouds are mainly located around the nodes of the respective lattices (glass or incommensurate), but that we have also displacements between different sites. 
This is the key signal for superfluidity. We also show in the same figure the $x-y$ static structure factors for both phases. As expected, the one for the glass does not show any 
Bragg peak and looks rather similar to $S(k)$ for a liquid at the same density. Instead, the triangular crystal shows a clear Bragg peak, but relatively small due to the  
delocalization of particles.  


In this work, we have studied the adsorption of H$_2$ on an amorphous substrate. 
To do so, we have used 
a set of coordinates that were supposed to model adequately an amorphous two-dimensional carbon material experimentally obtained \cite{e3}.  
We considered a both a flat substrate and a corrrugated one. 
Surprisingly, the results are quite similar in one important question: there is at least 
a region around 0.055 \AA$^{-2}$ for which we have an stable glass. We have also found that the superfluid 
density of that glass can be tiny or sizable, but not zero. This result is compatible with a recent calculation for the second layer of H$_2$ on 
graphite \cite{prb2022} that found a tiny supersolid density in a very thin density region. 
As in that work, we can adscribe the superfluidity to the relative low density of the glass at equilibrium. 
This prompts us to suggest that we can expect to find a superglass in     
a real disordered substrate similar to that of Ref. \onlinecite{e3}. In the worst case scenario, a superfluid density of 
the order of the one we found for the flat substrate can be detected using the perfected torsional oscillator technique used in Ref. \onlinecite{kim} for $^4$He on graphite. 

\begin{acknowledgments}
We acknowledge financial support from Ministerio de
Ciencia e Innovación MCIN/AEI/10.13039/501100011033 
(Spain) under Grants No. PID2020-113565GB-C22 and No.
PID2020-113565GB-C21, and from Junta de Andaluc\'{\i}a group
PAIDI-205. M.C.G. acknowledges funding from Fondo Europeo de Desarrollo Regional (FEDER) and Consejer\'{\i}a de
Econom\'{\i}a, Conocimiento, Empresas y Universidad de la Junta
de Andaluc\'{\i}a, en marco del programa operativo FEDER  Andaluc\'{\i}a
2014-2020. Objetivo específico 1.2.3. “Fomento y
generaci\'on de conocimiento frontera y de conocimiento orientado a los retos de la sociedad, desarrollo de tecnolog\'{\i}as
emergentes” under Grant No. UPO-1380159. Porcentaje de
cofinanciaci\'on FEDER 80\%. J.B. acknowledges financial support from Secretaria d’Universitats i Recerca del Departament
d’Empresa i Coneixement de la Generalitat de Catalunya, cofunded by the European Union Regional Development Fund
within the ERDF Operational Program of Catalunya (project
QuantumCat, Ref. No. 001-P-001644). We also acknowledge
the use of the C3UPO computer facilities at the Universidad
Pablo de Olavide.
\end{acknowledgments}

\bibliography{g3_boro2}

\end{document}